\documentclass[aps,prl,twocolumn,superscriptaddress]{revtex4}

\usepackage{graphicx}

\usepackage{color}
\usepackage{ulem}

\newcommand{\bfcavf}{Ba(Fe$_{0.955}$Co$_{0.045}$)$_2$As$_2$}
\newcommand{\bfcax}{Ba(Fe$_{1-x}$Co$_x$)$_2$As$_2$}
\newcommand{\bfcas}{Ba(Fe$_{0.94}$Co$_{0.06}$)$_2$As$_2$}
\newcommand{\bfa}{BaFe$_2$As$_2$}

\newcommand{\Angrz}{$\text{\AA}^{-1}$}

\newcommand{\tc}{$T_{C}$}

\newcommand{\chilon}{$\chi''_{long}$}
\newcommand{\chitil}{$\chi''_{t{-}in}$}
\newcommand{\chitol}{$\chi''_{t{-}out}$}

\begin{document}

%\preprint{}

\title{Suppression of low-energy longitudinal spin-excitations in Co-underdoped BaFe$_2$As$_2$}


\author{F. Wa\ss er}\email[e-mail: ]{wasser@ph2.uni-koeln.de}
\affiliation{$I\hspace{-.1em}I$. Physikalisches Institut,
Universit\"at zu K\"oln, Z\"ulpicher Str. 77, D-50937 K\"oln,
Germany}

\author{C.H. Lee}
\affiliation{National Institute of Advanced Industrial Science and Technology (AIST), Tsukuba, Ibaraki 305-8568, Japan}

\author{K. Kihou}
\affiliation{National Institute of Advanced Industrial Science and Technology (AIST), Tsukuba, Ibaraki 305-8568, Japan}

\author{P. Steffens}
\affiliation{Institut Laue Langevin, 71 avenue des Martyrs,
38000 Grenoble, France}

\author{K. Schmalzl}
	\affiliation{J\"ulich Centre for Neutron Science, Forschungszentrum J\"ulich GmbH, Outstation at Institut Laue-Langevin, 71 avenue des Martyrs,
38000 Grenoble, France}

\author{N. Qureshi}
\affiliation{$I\hspace{-.1em}I$. Physikalisches Institut,
Universit\"at zu K\"oln, Z\"ulpicher Str. 77, D-50937 K\"oln,
Germany}
\affiliation{Institut Laue Langevin,71 avenue des Martyrs,
38000 Grenoble, France}

\author{M. Braden}\email[e-mail: ]{braden@ph2.uni-koeln.de}
\affiliation{$I\hspace{-.1em}I$. Physikalisches Institut,
Universit\"at zu K\"oln, Z\"ulpicher Str. 77, D-50937 K\"oln,
Germany}

%\thanks{}


%\date{\today}

\begin{abstract}

Polarized inelastic neutron scattering experiments were performed to study magnetic excitations in the normal and superconducting phases of Co-underdoped BaFe$_2$As$_2$,
which exhibits coexistence of antiferromagnetic order and superconductivity. In the normal state the antiferromagnetic order results in broadened
spin gaps opening in all three spin directions that are reminiscent of the magnetic response in  pure antiferromagnetic BaFe$_2$As$_2$. In particular
longitudinal excitations exhibit a large gap. In the superconducting state we find two distinct resonance excitations,  which both
are anisotropic in spin-space, and which both do not appear in the longitudinal polarization channel. This behavior contrasts to previous polarized
neutron results on samples near optimum or higher doping. The gap in the longitudinal fluctuations arising from the antiferromagnetic order seems to
be sufficiently larger than twice the superconducting gap to suppress any interplay with the superconducting state. This suppressed low-energy weight of
longitudinal fluctuations can explain the reduced superconducting transition temperature in underdoped BaFe$_2$As$_2$ and indicates that the coexistence
of antiferromagnetism and superconductivity occurs locally.

\end{abstract}

\pacs{7*******}

%\keywords{}

\maketitle

\begin{figure}
\includegraphics[width=0.999\columnwidth]{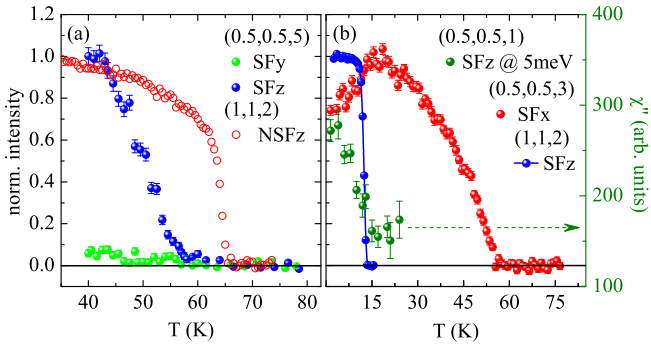}
\caption{\label{tdep} Temperature dependence of nuclear and magnetic Bragg scattering in \bfcavf ; the nuclear Bragg peak (1  1 2) exhibits
a sharp intensity increase at the structural phase transition, $T_S$$\sim$65\ K, while magnetic scattering at (0.5 0.5 5) appears at $T_N$$\sim$55\ K (a). The SC transition of our sample crystal was studied through the neutron depolarization yielding
a $T_C\sim$14\ K. The onset of superconductivity results in a suppression of the magnetic Bragg scattering at (0.5 0.5 3) as reported in \cite{20} (b)
and to an increase of inelastic scattering at (0.5 0.5 1) and 5\ meV.
}
\end{figure}
%=========================================================

Magnetism and superconductivity (SC) are closely connected in the FeAs-based superconductors with clear evidence for coexistence of SC and antiferromagnetic (AFM) order \cite{1,2}. The SC transition results in sizeable suppression of the ordered moment \cite{20,3}, but, so far there is incertitude whether the two phases coexist locally or whether there is phase separation. The emergence of magnetic resonance modes
in the SC phase yields another strong argument in favor of magnetic pairing \cite{4}, however, several inelastic neutron scattering (INS) experiments show that these resonance excitations are more complex than a simple triplet exciton \cite{lee2013,6,7,8,9}. Firstly, the resonance features exhibit a finite $c$ dispersion depending on the proximity to the AFM order \cite{lee2013}. Secondly,  polarized neutron experiments on optimum
Co-doped BaFe$_2$As$_2$ show that at least two resonance components appear at the SC transition, of which the lower is anisotropic \cite{6}. Such spin-space anisotropy must arise from the microscopic spin-orbit coupling
because dipolar effects can be safely neglected at an energy scale of several meV, but so far there is no consensus about the microscopic explanation of the
anisotropic resonances.
Polarized INS experiments on Ni  \cite{7} and K optimum doped BaFe$_2$As$_2$  \cite{8,9} reveal qualitatively the same double resonance excitations, and
also in K overdoped BaFe$_2$As$_2$  \cite{9} an anisotropic low-energy resonance remains visible ruling out that it just arises from quasi-static magnetic order.
In Co-doped NaFeAs \cite{10} and in Na-doped
BaFe$_2$As$_2$  \cite{11} the low-energy anisotropic features even dominate the extra magnetic scattering in the SC state. Various models were proposed to explain the
coexistence of the two resonance features. The low-energy resonance can be ascribed to quasistatic correlations while the isotropic resonance
at higher energy is explained by the usual triplet exciton from the SC phase \cite{12,12a}. In another approach the split resonance is attributed to
an orbital and band selective pairing which would retain some well defined orbital character in the low-energy resonance modes \cite{13}.

The observation of the anisotropic resonance features near optimum doping reflects the anisotropy gaps of the pure compound. In BaFe$_2$As$_2$ magnetic moments
order along the in-plane component of the propagation vector \cite{2}, which  corresponds to the [110] direction in the tetragonal lattice ($a$ in the orthorhombic notation).
The rotation of magnetic moments from this direction to $c$ represents the excitation with the lowest energy, while rotating moments in the plane
requires a larger energy \cite{14}. The small structural distortion accompanying the magnetic order thus results in unexpectedly strong in-plane anisotropy, which,
however, agrees with its strong signatures in charge and orbital properties \cite{15,16}. In this paper we will use the tetragonal notation
and denote the three directions of spin-space with respect to the ordered moment: longitudinal, $long$, transverse in-layer, $t{-}in$, and transverse out-of-layer, $t{-}out$. Note that the
choice of the scattering vector, mostly (0.5,0.5,$q_l$) selects the  $long$ and $t{-}out$ directions. The determination of $t{-}out$ (or $c$) as the second magnetically soft direction in AFM BaFe$_2$As$_2$ was recently corroborated by the observation of the spin-reorientation transition in Na-doped BaFe$_2$As$_2$, which
results in AFM ordered moment along $t{-}out$ (or $c$) direction \cite{17}.

%=========================================================
\begin{figure}
\includegraphics[width=0.9\columnwidth]{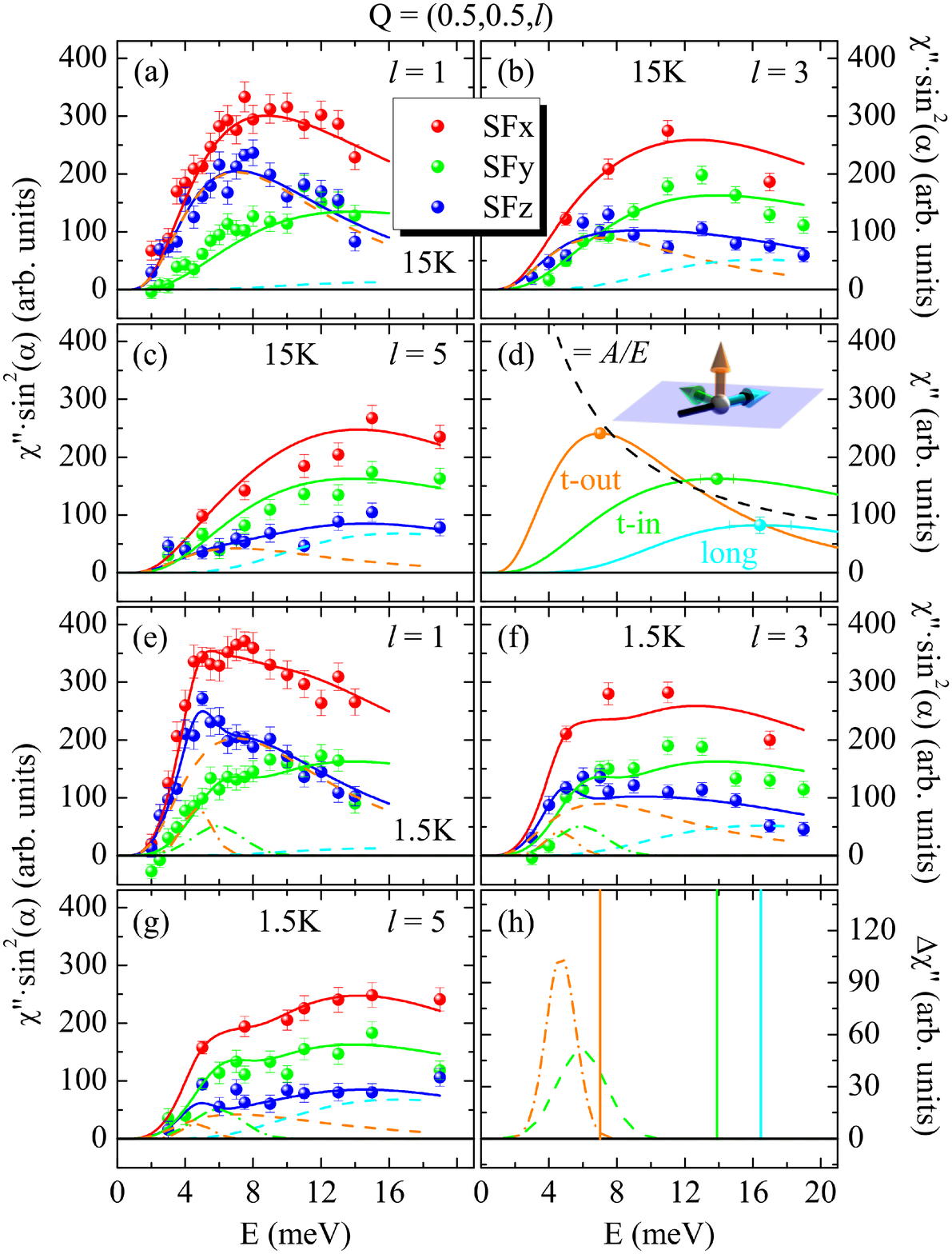}
\caption{\label{fig2} (a-c) Polarized INS intensity at (0.5 0.5 $l$) with $l$= 1, 3 and 5 in the different SF channels after subtracting the
background and correcting for the Bose and form factors. These data correspond to the dynamical susceptibilities
$\chi ''$ multiplied with the geometry factors. The nine curves are consistently fitted (lines) by the three susceptibilities
 $\chi ''_{long}$,  $\chi ''_{t{-}in}$ and $\chi ''_{t{-}out}$ each described by a single log-normal distribution. These
 individual susceptibilities are resumed in (d); their amplitude follows an 1/$E$ relation. (e-g) the same data in the SC phase
 at 1.5\ K where only two additional resonance components in  $\chi ''_{t{-}in}$ and $\chi ''_{t{-}out}$ are needed to again
 consistently describe all nine curves. These additional resonance contributions are shown in (h), where vertical lines denote the maximum response
 in the normal state (see (d)).  }
\end{figure}
%=========================================================

So far the polarized INS experiments on the spin-space anisotropy were performed for near to optimum or overdoped BaFe$_2$As$_2$
with strongly suppressed AFM correlations. Therefore, no sizeable spin gap can be expected, while the SC gaps are large.
For example in the 6\% Co-doped BaFe$_2$As$_2$  studied in reference  \cite{6} no long range magnetic order is observed while the
SC gap 2$\Delta$ amounts to 10\ meV \cite{18}, which is of the order of the gaps in the transverse magnons in pure BaFe$_2$As$_2$.
In this work we study 4.5 \% Co underdoped BaFe$_2$As$_2$ by polarized INS experiments. This composition still exhibits considerable
magnetic order, $T_N$=55\ K, while the superconductivity is reduced, $T_c$=14\ K. Therefore, the usual resonance can be expected near
4.3$\cdot$$k_BT_c\sim$5.2\ meV \cite{18}, which is well below the magnon gap of the pure compound. Many unpolarized neutron studies on almost the same
concentration were reported \cite{20,19,21,22,23}. The main result concerns the appearance of a broad resonance indeed near 4.5\ meV.
Our polarized experiments show that the magnetic response in the SC phase
for this underdoped material is fundamentally different from that for optimum doping. In underdoped BaFe$_2$As$_2$ there are two components of the resonance, but both
are anisotropic in spin space. In addition the longitudinal fluctuations are gapped by the AFM order and do not interplay with superconductivity.
% The gap in the longitudinal fluctuations gives a simple explanation for the reduced $T_c$ and it indicates that superconductivity and
%antiferromagnetism coexist locally.

Three single crystals of \bfcax\ with Co-doping of x=4.5\% and with a combined mass of 2.12\ g were grown by the
FeAs-flux method. The structural and magnetic transitions were observed at $T_S$$\sim$65\ K and
$T_N$$\sim$55\ K, respectively, by following Bragg peak intensities of nuclear and magnetic peaks, see Fig. 1.
The polarized inelastic neutron scattering experiments were performed on the IN20 and IN22 thermal triple
axis spectrometers at the Institut Laue-Langevin in Grenoble. Both spectrometers were operated with Heusler
monochromator and analyzer crystals, and a graphite filter was set between the sample and the analyzer in order to suppress higher
order contaminations. Most data was taken with the final wave vector of the neutron fixed to 2.662~\Angrz.
Experiments were performed with either the CRYOPAD device to assure zero magnetic field at the sample position or
with Helmholtz coils to guide the neutron polarization at the sample \cite{pol-neutr}. With the Helmholtz coils the guide field
was not varied in the SC state in order not to deteriorate the neutron polarization. The flipping ratio
measured on nuclear Bragg peaks amounted to 14 on IN20 and 16 on IN22. Polarized INS allows one to separate magnetic and nuclear
contributions and to split magnetic scattering according to the polarization direction of the magnetic signal \cite{pol-neutr}. In general INS only
senses magnetic signals that are polarized perpendicular to the scattering vector ${\bf Q}$ resulting in a geometry factor $\sin^2(\alpha)$ with $\alpha$ the angle between ${\bf Q}$ and the magnetic signal. With the longitudinal
polarization analysis this active part of the magnetic signal further splits. In the neutron spin-flip (SF) channel
one finds the part that is also perpendicular to the neutron polarization direction, while the neutron non-spin-flip (NSF) channel contains
the parallel polarization \cite{9}. We use the reference system with $x$ along the scattering vector ${\bf Q}$ (here all the magnetic scattering
contributes to the SFx channel), $z$ perpendicular to the scattering plane, and $y$ perpendicular to $x$ and $z$.
All experiments were performed in the [110]/[001] scattering plane.

%=========================================================
\begin{figure}
\includegraphics[width=0.999\columnwidth]{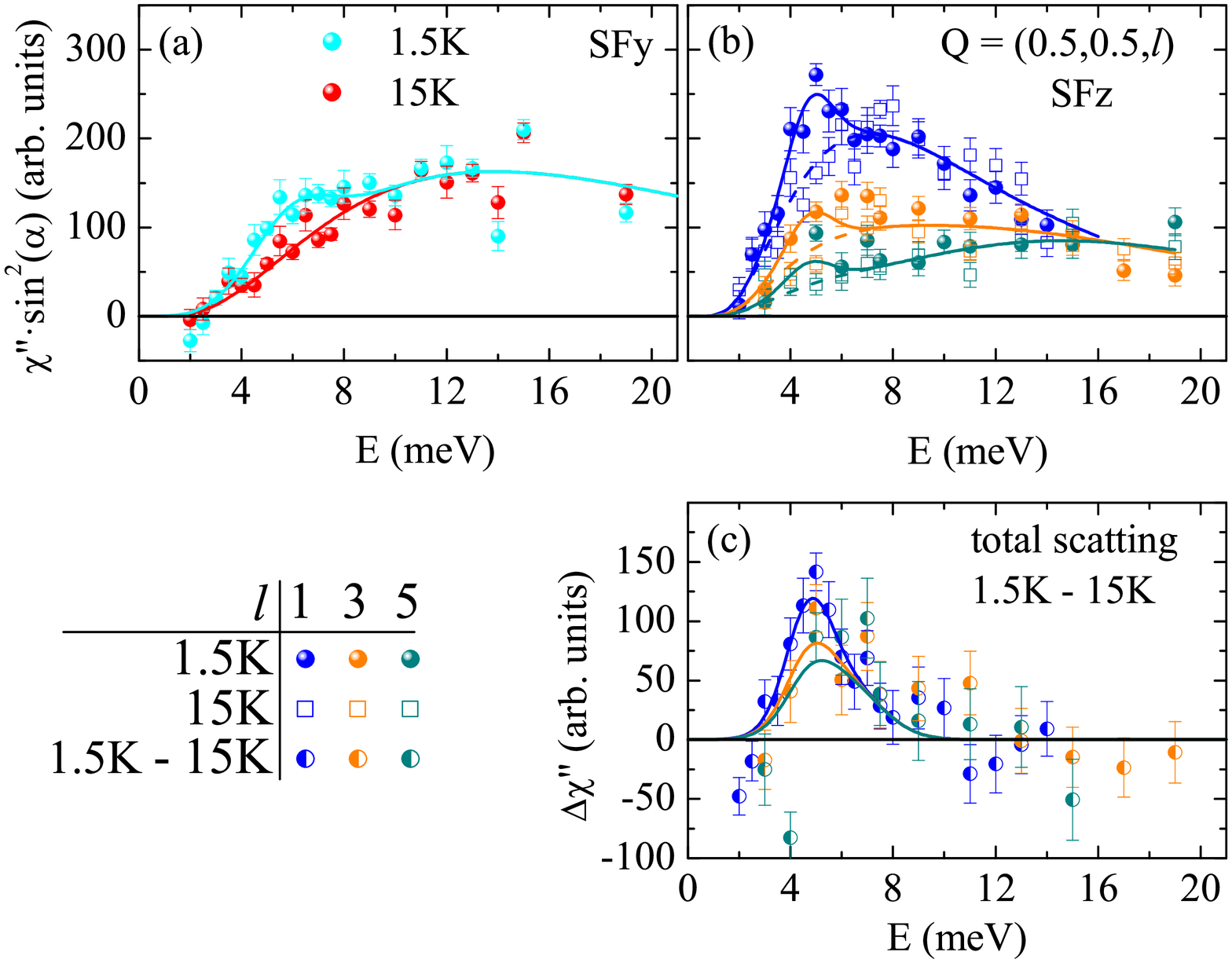}
\caption{\label{fig3}(a) Comparison of SFy intensities obtained above and below \tc ; since this channel only contains  $\chi ''_{t{-}in}$, the results obtained at $l$ = 1, 3 and 5 were added and lines correspond to the analysis presented in Fig. 1. (b) SFz scattering at different $l$ values with the analysis of Fig. 1. Note that the intensity enhancements in the SFy and SFz channels peak at different energies. (c) temperature difference of the total magnetic scattering as it is also observed in an unpolarized experiment.}
\end{figure}
%========================================================

Fig. 1 resumes the various transitions appearing in \bfcavf . 
Varying the guide fields in the SC state induces neutron depolarization through flux pinning, which  
can be used to determine the SC $T_c$=14\ K of the sample crystal in situ \cite{noteSC}, see Fig. 1 (b).
The emergence of orthorhombic domains results in a lower
local crystal quality and thus in a reduced extinction effect and enhanced Bragg intensity \cite{20,6}. Thereby we determine the
structural transition to $T_S\sim$65\ K. The magnetic transition occurs at $T_N\sim$55\ K as seen in the sharp rise of the
magnetic intensity. Both values are in perfect agreement with the reported phase diagrams \cite{2}. Close inspection of the
magnetic signals shows, that long-range intensity shows up in the SFz channel, which agrees with the in-layer alignment
of the moment. However, this anisotropy of SFz versus SFy persists in the diffuse scattering in the nematic phase between $T_N$ and $T_S$, see Fig. 1 (a). In this nematic phase
the fourfold spin-space symmetry is already broken and the magnetic diffuse signal only corresponds to longitudinal in-layer correlations, $long$ direction, that appear in SFz.
The SC transition, $T_C\sim$14\ K, is visible in the sharp drop of neutron polarization due to flux pinning, see Fig. 1(b), and
in the suppression of magnetic intensity by about 26\ \% in the SC state.
The latter was measured at the (0.5 0.5 3) magnetic Bragg peak. In addition there is an increase of
inelastic scattering at (0.5 0.5 1) and 5\ meV, which corresponds to appearance of the resonance mode, see Fig. 1 (b) and the discussion below.

Polarized INS scans at (0.5 0.5 $l$) with $l$ = 1, 3 and 5 are shown in Fig. 2. The ${\bf Q}$ values correspond to
AFM zone centers and the variation of the $l$ values allows one to separate the $long$ and $t{-}out$ directions. For small $l$ the $c$ or
$t{-}out$ direction is almost perpendicular to ${\bf Q}$ so that the SFz signal essentially corresponds
to the $t{-}out$ direction. In contrast, at large $l$ ${\bf Q}$ is nearly parallel to $c$ so that the SFz channel essentially contains
the $long$ direction. The $t{-}in$ contribution exclusively contributes to the SFy channel. The difference SFy+SFz-SFx gives a direct
determination of the background, and 2SFx-SFy-SFz the total magnetic scattering. Fig. 2 (a-d) show the signal in the SF channels
after subtracting the background. The data were corrected for higher-order contaminations of the monitor, {the Bose factor} and for the formfactor, so that they correspond to
the imaginary part of the susceptibilities multiplied with the geometry factors, $\sin^2(\alpha)$, described above.
The nine spectra can be consistently described by the three susceptibilities \chilon , \chitil and \chitol. Each of these is described
by an asymmetric log-normal distribution, $A_i exp(-\frac{(ln(E)-ln(\Gamma_i))^2}{\sigma_i^2})$, which well describes the spin-gap spectra with only three parameters. By a concomitant fit
of all spectra we obtain the three susceptibilities shown in Fig. 2(d). The magnetic response of \bfcavf \ can be well understood as
the spin-wave-like response of a magnetically ordered material with disorder  \cite{14,20}. In the transversal channels $t{-}in$ and $t{-}out$, the well-defined
spin gaps of the pure material at
{18.9} and
{11.6\ meV} are renormalized to maxima at
{13.9} and
{7.0\ meV}. The asymmetric shape of these signals is
well described by the log-normal distribution; it results from the instrumental resolution and some disorder induced broadening.
Also in the $long$ channel we find a pseudogap at $\sim$16\ meV. In view of the report of a longitudinal gap in pure
\bfa \ of only 24\ meV  \cite{24},  this would indicate a surprisingly small renormalization by 4.5\ \% Co doping. In addition the strength of the longitudinal signal
is comparable to that of the two transversal directions in the underdoped material, while much smaller longitudinal weight is reported in the pure crystal \cite{24}.
This sheds additional doubts on the interpretation of the very weak signal
in pure \bfa \ as the longitudinal mode and supports an alternative  two-magnon explanation  \cite{25}.

%=========================================================
\begin{figure}
\includegraphics[width=0.9\columnwidth]{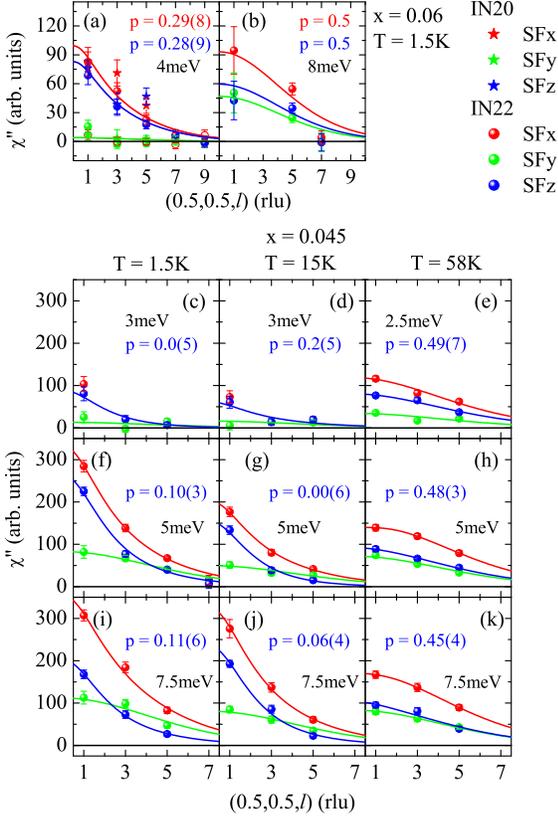}
\caption{\label{fig4} $l$ dependence of the scattering in various SF channels which allows one
to separate the  $\chi ''_{long}$ and $\chi ''_{t{-}out}$ components in the SFz channel. The parameter $p$
describes the contribution of $\chi ''_{t{-}out}$ required to fit the $l$ dependence.
Panel (a) shows
the low temperature results for the 4\ meV signal in optimum doped \bfcas , which has no $t{-}in$ component but consists
of 29\ \% of $long$ signal. In contrast, the isotropic resonance at $\sim$8\ meV appears equally strong, $p$=0.5, in the three directions (b).
Panels (c)-(k) present the same analysis  in \bfcavf \ for energies of 3, 5 and 7.5\ meV, when not labelled differently, at
T=1.5\ K in the SC state, at 15\ K in the non-SC AFM phase, and at 58\ K in the paramagnetic phase.}
\end{figure}
%=========================================================

In Fig. 2 (e-g) we present the same analysis of the magnetic scattering in the SC state. We can describe the total
response at the three studied scattering vectors by adding two resonance features, one in $\chi ''_{t{-}out}$ and one in
$\chi ''_{t{-}in}$, to the fixed normal state susceptibilities determined at 15\ K.
That there are indeed two distinct resonance modes can be further seen in the SFy and SFz channels shown
in Fig. 3 (a) and (b), respectively. The SFy only senses the $\chi ''_{t{-}in}$ with full geometry factor; therefore, we
summed the data taken at the different $l$ values. This $\chi ''_{t{-}in}$ resonance peaks at 4.7\ meV. In contrast,
$\chi ''_{t{-}out}$ and  $\chi ''_{long}$ contribute to the SFz channel with varying geometry factors (that always sum
to one). However, the  $\chi ''_{long}$ remains fully suppressed at energies below $\sim$8\ meV, therefore the additional
resonance signal in the SFz channels always stems from $\chi ''_{t{-}in}$ and it peaks at {5.9\ meV}.
These resonance energies were obtained by a simultaneous fit of the the spectra at 1.5\ K.
That there are two distinct resonance contributions can also be seen when summing up all magnetic scattering at the three $l$ values, see Fig. 3 (c), which resembles the
unpolarized data taken previously  \cite{20,19,21,22,23}. Our analysis can be corroborated by fitting the $l$ dependence of the
magnetic signals shown in Fig. 4. A larger $l$ favors the observation of the $\chi ''_{long}$ on the dispense
of $\chi ''_{t{-}out}$ in the SFz channel. Therefore, we may determine the ratio $p$ of the $long$ component defined as $ p=\frac{\chi ''_{long}}{\chi ''_{long}+\chi ''_{t{-}out}}$. At $T$=58\ K
above the N\'eel temperature all signals are isotropic, $p$=0.5, within the error bars, and there is little difference between
the SFz and SFy channels (except at 2.5\ meV sensing critical scattering). However, at 15\ K in the AFM phase and at 1.5\ K in the
SC and AFM phase we find an insignificant $long$ component in the SFz channel, $p\sim$0, in agreement with the conclusion
that longitudinal excitations are gapped due to the significant ordered moment. Fig. 4 (a) shows the same $l$ analysis for
the additional low-energy resonance signal in optimum-Co-doped \bfa (x=0.06). The slow reduction of the SFz signal with $l$ indicates that
there is a significant, $p$=0.28, longitudinal component in this resonance mode at optimum doping;
similar results were reported for optimum Ni doped  \cite{7} and
K overdoped \cite{9} BaFe$_2$As$_2$.

The anisotropy of the resonance excitations in underdoped \bfcavf \ is thus fundamentally different from the near-optimum or
overdoped \bfa \ crystals studied previously by polarized INS experiments and excludes a recently proposed explanation \cite{note}. There is no isotropic feature in this underdoped
material which can be associated with the usual spin triplet exciton. Instead there are two polarized resonance modes appearing
in the $\chi ''_{t{-}out}$ and $\chi ''_{t{-}in}$ channels. The reason for these fully anisotropic resonances seems to consist
in the gap that opens in $\chi ''_{long}$ due to the AFM ordering. Although this longitudinal gap is  renormalized from the
expected one in pure \bfa \ (considering the optical studies  \cite{26,26a}) it still is considerably larger than twice the SC
gap 2$\Delta$$\sim$8\ meV determined in ARPES experiments  \cite{18}.  Therefore, longitudinal fluctuations cannot interplay with the superconductivity in this underdoped material and they cannot contribute to a lowering of exchange energy  \cite{4}. This yields a simple explanation for the reduced SC transition temperature.
The fact that we do not see longitudinal excitations in the SC state in contrast to the optimum or overdoped
samples, furthermore, indicates that superconductivity and AFM ordering coexist locally.

In conclusion we have studied the magnetic response in underdoped \bfcavf \ by polarized INS experiments that reveal a different
behavior compared to previously studied compounds with a larger amount of doping. The significant ordered moment results in
sizeable pseudogaps opening in the magnetic excitations along all three directions. In the longitudinal channel this gap clearly exceeds twice the SC
one, therefore longitudinal fluctuations remain unaffected by the SC transition. In the SC state
two resonance components can be separated that are both anisotropic in spin space, one appearing in $\chi ''_{t{-}out}$  the
other in $\chi ''_{t{-}in}$. The full suppression of low-energy longitudinal fluctuations in the SC state for 4.5 \% Co and the observation
of sizeable longitudinal excitations for optimum doping strongly suggest
that superconductivity and AFM ordering coexist microscopically in Co underdoped BaFe$_2$As$_2$.

\begin{acknowledgments}
This study was supported by a Grant-in-Aid for Scientific Research
B (No. 24340090) from the Japan Society for the Promotion of
Science and by the Deutsche Forschungsgemeinschaft through the
Priority Programme SPP1458 (Grant No. BR2211/1-1).

\end{acknowledgments}

%\stophere


\begin{thebibliography}{32}
\bibitem{noteSC} Fig. 1 (b) shows the {(1 1 2)} nuclear Bragg peak intensity measured in the SFz channel after rotating the guide field
generated with the Helmholtz coils from $x$ to $z$ direction in the SC phase. The flux pinning in the SC state prohibits the proper rotation of the guide field and reduces the neutron polarization yielding sizeable nuclear intensity in the SFz channel. When heating
above $T_c$ \ the loss of SC flux pinning restores the polarization and therefore reduces the
Bragg intensity in SFz to that corresponding to the finite flipping ratio.
\bibitem{note} The two resonance excitations have recently been  attributed to an isotropic two-dimensional exciton  and a spin-wave like feature
stemming from the ${t{-}out}$ anisotropy mode that sharpens upon opening of the SC gap  \cite{18}. This picture cannot explain
the two anisotropic resonances in our underdoped crystal, nor the fact that for optimum doping the low-energy resonance
also possesses a longitudinal component, see Fig. 4 (a). Furthermore, recent studies of Na-doped \bfa , which exhibits a spin reorientation,
disagree with such theory \cite{11}.
\end{thebibliography}
\end{document}